\documentclass[twocolumn,aps,prl,floatfix,showpacs]{revtex4}

\usepackage{graphicx}
\usepackage{dcolumn}
\usepackage{bm}

\begin{document}

\title{Novel Spin-texture on the warped Dirac-cone surface states in
topological insulators }

\author{Susmita Basak$^1$, Hsin Lin$^1$, L.A. Wray$^{2,3}$,
S.-Y. Xu$^{2,3}$, L. Fu$^4$, M.Z. Hasan$^{2,3}$, A. Bansil$^1$}

\affiliation{$^1$Department of Physics, Northeastern University, Boston,
Massachusetts 02115, USA}

\affiliation{$^2$Joseph Henry Laboratories of Physics, Princeton
University, Princeton, New Jersey 08544, USA}

\affiliation{$^3$Princeton Center for Complex Materials, Princeton
University, Princeton, New Jersey 08544, USA}

\affiliation{$^4$Department of Physics, Harvard University, Cambridge,
Massachusetts 02138, USA}

\begin{abstract}

We have investigated the nature of surface states in the Bi$_2$Te$_3$ family of
three-dimensional topological insulators using first-principles calculations
as well as model Hamiltonians. When the surface Dirac cone is warped due to
Dresselhaus spin-orbit coupling in rhombohedral structures, the spin acquires
a finite out-of-plane component. We predict a novel in-plane spin-texture of
the warped surface Dirac cone with spins {\em not} perpendicular to the electron momentum.
Our $k\cdot p$ model calculation  reveals that this novel in-plane spin-texture requires high order Dresselhaus spin-orbit coupling terms.

\end{abstract}

\maketitle

Topological insulators (TI) \cite{hasankane,fukane,davidbisb,bitedavid,moore1,fu1,roy,hgte,konig}  realize a novel state of quantum matter that is distinguished by topological invariants of bulk band structure rather than spontaneously broken symmetries. Its material realization in 2D artificial HgTe-quantum wells \cite{hgte,roy,konig} and 3D Bi-based compounds \cite{bise,davidprl,roushan,bisezhang} has led to a surge of interest in novel topological physics throughout the world-wide condensed matter community.
The surfaces of 3D topological insulators or the interface between two materials with distinct topological invariants host metallic surface/interface states.
Electrons on the surface of an ideal topological insulator have an energy-momentum relationship in the shape of a Dirac cone.  In the presence of such a single-Dirac-cone
surface or interface band, a  number of exotic quantum phenomena have been predicted\cite{majorana,fuproximity,leek,wolf}.
Recent spin-resolved photoemission measurements reveal that the spins of the electrons on the surface Dirac cone are locked with their momenta, giving rise to helical Dirac fermions without spin degeneracy \cite{davidbisb,bise,bite,bitedavid,nishidearpes,bitespin}. This one-to-one locking of the electron spin to the momentum comes from a combination of strong spin-orbit interaction and inversion symmetry breaking at the surface. Such spin-texture on the surface Dirac cone leads to anti-localization properties and plays a central role for exotic quantum phenomena.

Recent angle-resolved photoemission (ARPES) experiments have observed the surface Dirac cone in a number of topological insulators \cite{bitedavid,bite,davidprl}. In particular, Bi$_2$Te$_3$  exhibits a Dirac cone with significant hexagonal warping\cite{bitedavid,bite,davidprl,fuprl}, i.e. the shape of Fermi surface (FS) or constant energy contour evolves from an ideal circle to a hexagon to a snowflake-like shape with increasing energy. The hexagonal distortion of the cone does not change the
Berry phase quantifying its topological invariant,
consistent with its topological order \cite{bitedavid,davidprl}.
Fu \cite{fuprl} proposed that Dresselhaus spin-orbit coupling can lead to hexagonal warping of the Dirac cone in rhombohedral structures.
The spins acquire finite out-of-the-plane components through the orbital channel to conserve the net value of the Berry's phase. The resulting finite value of the out-of-plane component opens up new possibilities for unusual phenomena such as the enhancement of interference patterns around crystal defects and a magnetically ordered exotic surface.

In this Letter, we investigate the nature of surface states in the Bi$_2$Te$_3$ family of
three-dimensional topological insulators using first-principles calculations
as well as model Hamiltonians. We reproduce a warped surface Dirac cone with
finite out-of-plane spin component, and further predict a novel in-plane spin-texture
with spins {\em not} perpendicular to the electron momentum.
We develop a $k\cdot p$ model to describe this novel in-plane spin-texture which requires high order Dresselhaus spin-orbit coupling terms, which has not been previously proposed.
This is a generic property for surface states in strong spin-orbit coupling materials, not limited to topological insulators.

\begin{figure*}
\includegraphics[width=1\textwidth]{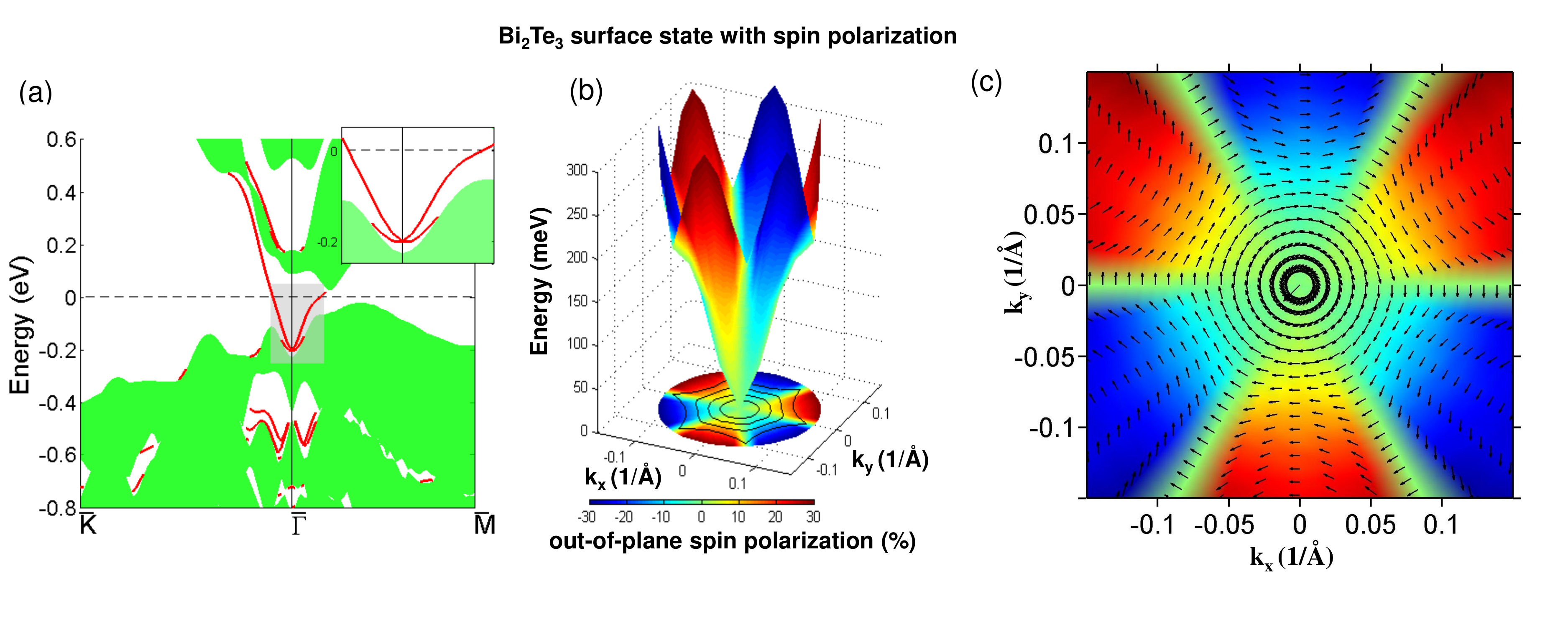}
\caption{\label{fig:sketch} (Color Online)
(a) Electronic structure of Bi$_2$Te$_3$. The surface
states are shown as red lines, projected bulk bands by shaded area. The inset is the area shaded in gray.
(b) Energy dependence of Bi$_2$Te$_3$ surface Dirac cone illustrating hexagonal warping. The magnitude of the out-of-plane spin component is indicated by the colors -- see color bar.
(c) In-plane spin direction of the surface Dirac cone.}
\end{figure*}

Bi$_2$Te$_3$ is a semiconductor with a rhombohedral crystal structure (space group R$\bar{3}$m).
The unit cell contains five atoms, with quintuple layers ordered in Te(1)-Bi-Te(2)-Bi-Te(1) sequence.
Due to the weak bonds between two  Te(1) atoms adjacent quintuple layers, the (111) surface with Te(1) termination can be easily obtained and is the surface usually studied in experiments.
To obtain the surface states, we use a symmetric 30-atomic-layer slab in a hexagonal unit cell.
The theoretical band calculations are performed with
the full-potential linear augmented plane wave (LAPW) method using the WIEN2K package \cite{wien2k}. The generalized
gradient approximation (GGA) of Perdew, Burke and Ernzerhof \cite{PBE96} was used to describe
the exchange-correlation potential. Spin-orbit coupling was included as a second variational
step using scalar-relativistic eigenfunctions as basis.
After self-consistent charges and potentials are obtained in the symmetric slab,
we add an extra potential on one side of the surface for band computations to remove the degeneracy that
arises from the two opposite sides of the surface.

We show the computed surface bands (red) as well as the projected bulk band structure (green) in Fig.~1(a) along high symmetry directions in the surface Brillouin zone (BZ).
Note that the lower Dirac cone has a particularly distorted shape (inset in Fig.~1(a)).
The insulating phase of the bulk is predicted with a 45 meV indirect gap,
while the experimental value is 165 meV \cite{bitedavid}. This underestimation of band gap is a typical problem of GGA.
A metallic surface band arises due to the strong spin-orbit coupling. Between the $\bar{\Gamma}$-point and the $\bar{M}$-point, an odd number of surface bands cross the Fermi level ($E_F$). This is the hallmark of the topological insulator. Topological non-triviality can also be proved by calculating the time-reversal invariant $Z_2$ value using wave-function parity analysis, proposed by Fu and Kane \cite{fukane}. The value of $Z_2$ is found to be -1 due to a single band inversion at the $\Gamma$-point.

The topological surface states form a 2+1 dimensional Dirac-cone-like band dispersion in  energy-momentum space as shown in Fig.~1(b). The shapes of the energy contours for the upper Dirac-cone (projected onto the $E=0$ plane in Fig.~1(b)) are circles for the low energy region, but become hexagonal at $\sim$150 meV above the Dirac point. The hexagonal warping becomes more pronounced at higher energy where a snowflake-like shape is obtained around 200 meV. These surface states have been observed in ARPES experiments  \cite{bitedavid,bite,davidprl,fuprl} and the hexagonal warping effect was described in a $k\cdot p$ model by Fu\cite{fuprl}. Recent Fourier-transform scanning tunneling microscope (FT-STM) experiments have shown the nontrivial interference pattern due to nesting of the hexagonally warped energy contours\cite{vidya}.
When the Dirac cone is warped, it must carry a finite out-of-plane spin component.
The out-of-plane spin component has three-fold symmetry with up and down spin alternation around a circle, whereas the band dispersion displays six-fold symmetry. Since the mirror plane is along the $\bar{\Gamma}-\bar{M}$ direction (chosen as $x$-direction), the out-of-plane spin component has to be zero along this direction as shown in Fig.~1(b).  The in-plane spin texture in Fig.~1(c) demonstrates a new phenomenon, non-orthogonality between spin and momentum. The helical nature of the spin is preserved at low energies near the Dirac point, where the spins remain perpendicular to $k$. When $E$ is increased, a significant departure from this ideal helical behavior is observed, although the spin and momentum remain orthogonal to each other along the high symmetry directions $\Gamma$-M and $\Gamma$-K. Similar features were also observed in recent photoemission experiments on $\rm Bi_2Se_3$ \cite{gedik} and it was noted that the existing $k\cdot p$ theory could not explain the non-orthogonality between electron spin and momentum. To understand the origin of this non-trivial behavior of the spins, we study the surface states of $\rm Bi_2Te_3$ using a $k\cdot p$ model with higher order corrections in the following paragraph.

In our formalism, the effective Hamiltonian for the surface bands is expanded up to fifth order in $\vec k$, taking the following form,
\begin{eqnarray}
H(k) &=& E_0(k)+v_k(k_x\sigma_y- k_y\sigma_x)
+{\lambda_k\over{2}}(k_{+}^3+k_{-}^3)\sigma_z
\nonumber \\
&+& i\zeta (k_{+}^5\sigma_+ - k_{-}^5\sigma_-),
\label{hamil}
\end{eqnarray}
where $E_0(k)=k^2/2m^*$, $k_\pm=k_x \pm ik_y$, $\sigma_\pm=\sigma_x \pm i\sigma_y$ and the $\sigma_i$ are Pauli matrices.  The form of $H(k)$ is suitable for describing the [111] surface band structure near the $\Gamma$ point in the surface Brillouin zone of  $\rm Bi_2Te_3$ family TIs and
it is invariant under time-reversal and $C_{3v}$ symmetries.  $E_0(k)$ generates particle-hole asymmetry and the term, $H_0(k)=v_k(k_x\sigma_y- k_y\sigma_x)$, describes an isotropic 2D helical Dirac fermion. The Dirac velocity, $v_k=v(1+\alpha k^2+\beta k^4)$, where $v$ is the Fermi velocity, contains a fourth order correction term. More importantly, the $k^3$-term, $H_w={\lambda_k\over{2}}(k_{+}^3+k_{-}^3)\sigma_z $, leads to the hexagonal warping of the Fermi surface. The hexagonal warping parameter also has a second order correction term, $\lambda_k = \lambda(1+\gamma k^2)$. The last term in Eq. \ref{hamil} describes fifth-order spin-orbit coupling at the surface of rhombohedral crystal systems, and, to the best of our knowledge, it has not been reported before. The purpose of introducing this term is to explain the non-orthogonality of the spins, which cannot be reproduced by the $k^3$-term. We determine the parameters by fitting the model spin directions to the first principles results. The resulting values of the parameters are recorded in table \ref{para}.

\begin{table}[h]
\centering
\caption{Parameters used in $k\cdot p$ calculations.}
\begin{tabular}{|c|c|}
\hline
$\rm Parameters$ &  $\rm Bi_2Te_3$   \\
\hline
$m^*~(\rm in~  eV^{-1}\cdot\AA^{-2})$  & 35.21 \\
$v ~(\rm in~ eV\cdot\AA)$  & -0.0005 \\
$\alpha~(\rm in~  eV\cdot\AA^2)$  & 14.82 \\
$\lambda~(\rm in~  eV\cdot\AA^3)$   & -0.04 \\
$\beta~(\rm in~  eV\cdot\AA^5)$  &-1.26 \\
$\gamma~(\rm in~  eV\cdot\AA^5)$   & -0.001 \\
$\zeta~(\rm in~  eV\cdot\AA^5)$   & 0.35 \\
	\hline
\end{tabular}
\label{para}
\end{table}

\begin{figure}[htp]
\centering
\includegraphics[width=1\textwidth]{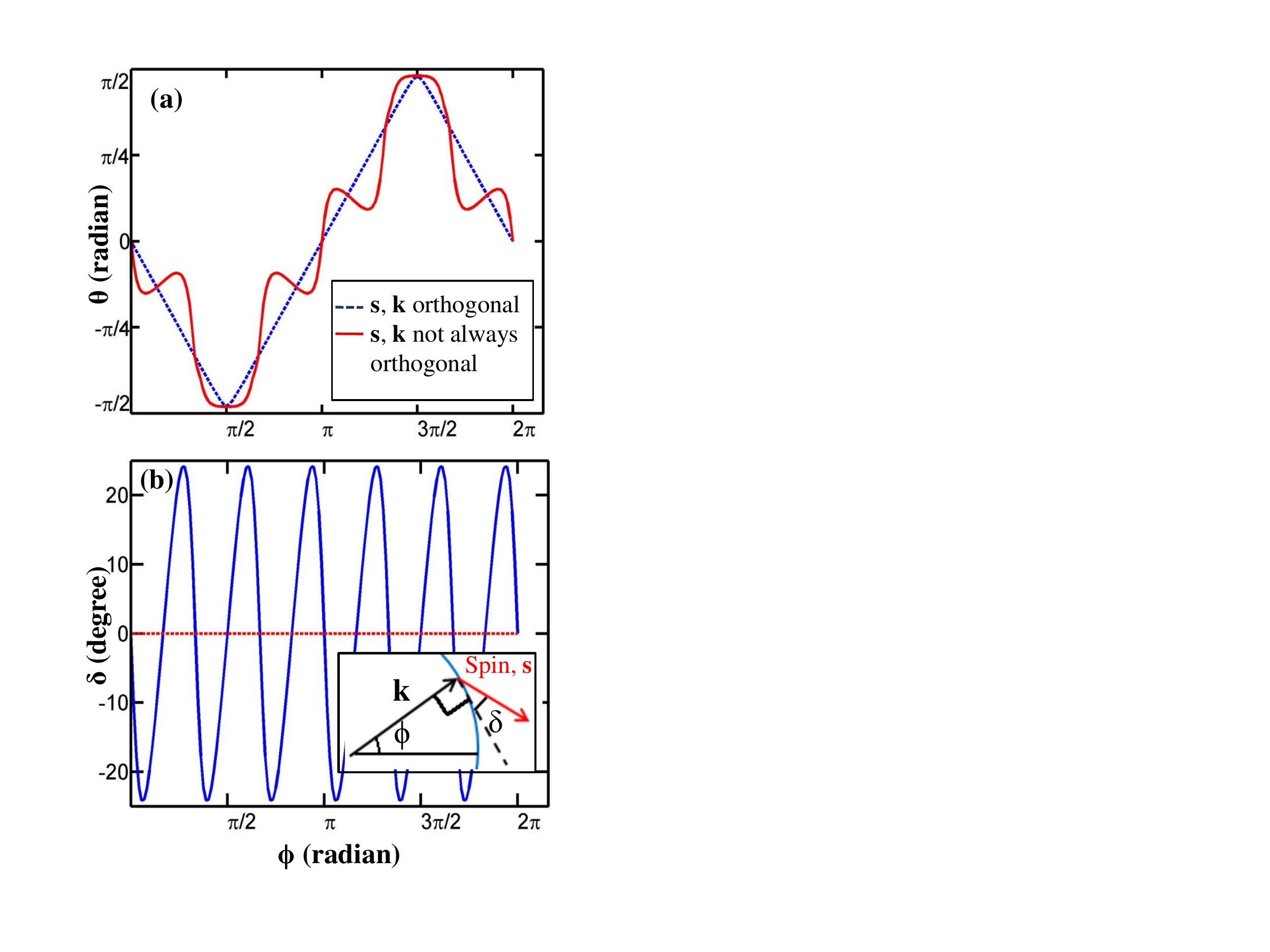}
\caption{ Quantitative analysis of non-orthogonality of electron spin and momentum on the surface Dirac cone of  $\rm Bi_2Te_3$. (a) Angle  $\theta$ between spin $\textbf s$, obtained from  $k\cdot p$ theory and $k_x$ is plotted against azimuthal angle, $\phi$ at an energy $-30$ meV. The blue dashed curve and solid red curve represent ideal and non-ideal helical behavior, respectively. (b) Angle of deviation, $\delta$ (inset) vs. $\phi$ plot.}
\label{angle}
\end{figure}

The in-plane spin polarization obtained from our model calculations agrees with the first-principles result by  capturing the non-orthogonality between electron spin and momentum. For a quantitative analysis, in Fig. \ref{angle} (a) we plot the angle ($\theta$) between the spin direction obtained from $k\cdot p$ theory and $k_x$ direction as a function of the azimuthal angle, $\phi$ at an energy $-30$ meV. For ideally helical spins, $\theta$ would follow the blue dashed line, but our calculated spins follows the red curve displaying a complicated pattern of spin canting.  The calculated angle is equal to the ideal angle only at high symmetry directions where the two curves intersect.
In Fig. \ref{angle} (b) we highlight the deviation of the angle between spin and momentum from orthogonality by plotting the angle of deviation, $\delta$ as a function of $\phi$.  $\delta$ (defined in the inset) oscillates about $0$ as we move along a constant energy contour in the higher $k$-region, where the $k^5$-term dominates making non-orthogonal feature more prominent. Here also we find that $\delta$ is exactly equal to $0$ along $\Gamma$-M and $\Gamma$-K directions, indicating that electron spin and momentum remain orthogonal along high symmetry directions in $k$-space.

Figure~3 shows the surface state dispersions obtained from the $k\cdot p$ model along the $\Gamma$-M (b) and $\Gamma$-K directions (d).  The dispersion of the surface state starts out linear in momentum $k$, as expected for a massless Dirac cone, but both lower and upper branches of the dispersion curve rapidly deviate from linearity, becoming convex functions of $k$, similar to the GGA results shown in Fig. 3 (a) and (c).
Both STM and ARPES studies of $\rm Bi_2Se_3$ capture similar behavior of the surface states \cite{hanaguri,bise}.

\begin{figure}[htp]
\centering
\includegraphics[width=1\textwidth]{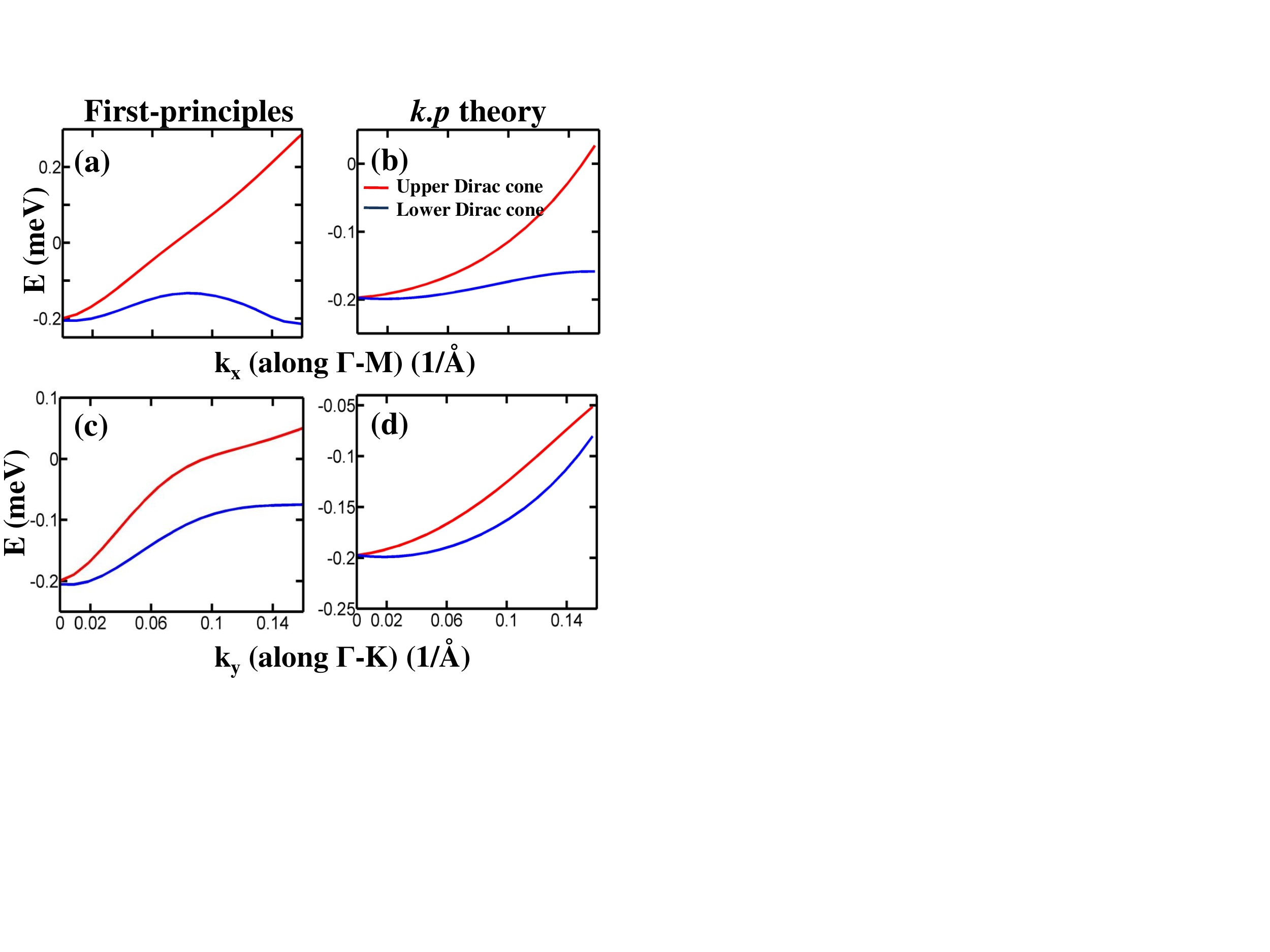}
\caption{ Dispersion of the surface states of $\rm Bi_2Te_3$ from (a),(c) first-principles calculations and (b),(d) $k\cdot p$ model.}
\label{band}
\end{figure}

In conclusion, we have shown that the spin of an electron on the surface of a strongly spin-orbit coupled material does not have to be orthogonal to the electron momentum due to higher order Dresselhaus spin-orbit coupling.
 This generic property should exist in both topological insulators and non-topological materials (metals or insulators). For example, pure bismuth is topologically trivial and may have novel spin-texture.
The non-orthogonality of spin and momentum should be observable on the warped surface Dirac cone in the topological insulator Bi$_2$Te$_3$ where the spin-texture is crucial for a number of exotic quantum phenomena.

The work at Northeastern and Princeton is supported by the
Basic Energy Sciences, US Department of Energy (DE-FG02-07ER46352,
DE-FG-02-05ER46200 and AC03-76SF00098), and benefited from the allocation
of supercomputer time at NERSC and Northeastern University's Advanced
Scientific Computation Center (ASCC). Support of the A. P. Sloan Foundation
(LAW and MZH) and the Harvard Society of Fellows (LF) is acknowledged.

\end{document}